\documentclass[fleqn,twoside,11pt]{article}

\usepackage{amssymb}
\usepackage{graphicx}
\usepackage{amsmath}

\begin{document}


\begin{center}

{\Large\bf Broken SU(3) Flavor Symmetry}

{\Large\bf  and Tribimaximal Neutrino Mixing 
}\footnote{
Contributed paper to XXIII International Symposium on Lepton and Photon
Interactions at High Energy (Lepton-Photon 2007), Aug 13-18, 2007, 
Daegu, Korea.
}

\vspace{3mm}
{\bf Yoshio Koide}

{\it IHERP, Osaka University, \\
1-16 Machikaneyama, Toyonaka, Osaka 560-0043, Japan\\
E-mail address: koide@het.phys.sci.osaka-u.ac.jp}

\date{\today}
\end{center}

\begin{abstract}
Recent work on a lepton mass matrix model based on an SU(3) 
flavor symmetry which is broken into S$_4$ is reviewed.
The flavor structures of the masses and mixing are caused by VEVs of
SU(2)$_L$-singlet scalars $\phi$ which are nonets 
({\bf 8}+{\bf 1}) of the SU(3) flavor symmetry, and which are broken 
into ${\bf 2}+{\bf 3}+{\bf 3}'$ and ${\bf 1}$ of S$_4$.
If we require the invariance under the transformation
$(\phi^{(8)},\phi^{(1)}) \rightarrow (-\phi^{(8)},+\phi^{(1)})$ 
for the superpotential of the nonet field $\phi^{(8+1)}$, the model 
leads to a beautiful relation for the charged lepton masses.
The observed tribimaximal neutrino mixing is understood by
assuming two SU(3) singlet right-handed neutrinos $\nu_R^{(\pm)}$
and an SU(3) triplet scalar $\chi$.
\end{abstract}

\vspace{3mm}


{\large\bf 1 \ Introduction}

The observed mass spectra and mixings of the fundamental particles will
provide promising clues to unified understanding of the quarks and leptons,
especially, to the understanding of the ``flavor".
In the present paper, we notice the following observed characteristic
features in the lepton sector \cite{PDG06}:

\noindent
(i) The observed charged lepton masses $(m_e, m_\mu, m_\tau)$
satisfy the relation \cite{Koidemass,Koide90}
$$
m_e+m_\mu+m_\tau=\frac{2}{3}(\sqrt{m_e}+\sqrt{m_\mu}+\sqrt{m_\tau})^2 ,
\eqno(1.1)
$$
with remarkable precision; 

\noindent
(ii) The observed neutrino mixing $U_{\nu}$ is approximately given by
 the so-called tribimaximal mixing \cite{tribi}
$$
U_{TB}=\left(
\begin{array}{ccc}
\frac{2}{\sqrt6} & \frac{1}{\sqrt3} & 0 \\
-\frac{1}{\sqrt6} & \frac{1}{\sqrt3} & -\frac{1}{\sqrt2} \\
-\frac{1}{\sqrt6} & \frac{1}{\sqrt3} & \frac{1}{\sqrt2}
\end{array} \right) ,
\eqno(1.2)
$$
which suggests that the mixing can be
described by Clebsh-Gordan-like coefficients. 
Therefore, for a start, in the present paper, we investigate
the lepton masses and mixings.
The purpose of the present paper is to review a recent 
attempt \cite{Koide0705} to investigate the mass relation (1.1) and the
tribimaximal mixing.

In order to understand the relation (1.1), for example, we assume
that there are three scalars $\phi_i$ ($i=1,2,3$), and the values of
the charged lepton masses $m_{ei}$ are proportional to the square of
the vacuum expectation values (VEVs) $v_i =\langle\phi_i\rangle$
of the scalars $\phi_i$, $m_{ei} = k v_i^2$ (in the 
Ref.\cite{Koide90,KF96,KT96}, for instance, a seesaw type model 
$(M_e)_{ij} =  \delta_{ij} v_i (M_E)^{-1} v_j$ has been assumed). 
We define singlet $\phi_\sigma$ and doublet $(\phi_\pi, \phi_\eta)$ 
of a permutation symmetry S$_3$ \cite{S3} by 
$$
\left(
\begin{array}{c}
\phi_\pi \\
\phi_\eta \\
\phi_\sigma 
\end{array}
\right) =
\left(
\begin{array}{ccc}
0 & -\frac{1}{\sqrt2} & \frac{1}{\sqrt2} \\
\frac{2}{\sqrt6} & -\frac{1}{\sqrt6} & -\frac{1}{\sqrt6} \\
\frac{1}{\sqrt3} & \frac{1}{\sqrt3} & \frac{1}{\sqrt3} 
\end{array} \right)
\left(
\begin{array}{c}
\phi_1 \\
\phi_2 \\
\phi_3
\end{array} \right) ,
\eqno(1.3)
$$
from the three objects $(\phi_1, \phi_2, \phi_3)$, 
and we consider the following S$_3$ invariant scalar potential 
$V(\phi)$ \cite{Koide90,Koide99,Koide06}:
$$
V(\phi) = m^2 (\phi_\pi^2 +\phi_\eta^2 +\phi_\sigma^2)
+\lambda_1 (\phi_\pi^2 +\phi_\eta^2 +\phi_\sigma^2)^2
+\lambda_2 \phi_\sigma^2 (\phi_\pi^2 + \phi_\eta^2).
\eqno(1.4)
$$
The minimizing condition of the potential (1.4) leads to
the relation
$$
v_\pi^2+v_\eta^2=v_\sigma^2 .
\eqno(1.5)
$$
The relation (1.5) means 
$$
v_1^2+v_2^2+v_3^2=\frac{2}{3}(v_1+v_2+v_3)^2 ,
\eqno(1.6)
$$
because 
$$
v_1^2+v_2^2+v_3^2=v_\pi^2+v_\eta^2+v_\sigma^2=2v_\sigma^2
 = 2 \left( \frac{v_1+v_2+v_3}{\sqrt{3}}\right)^2 .
\eqno(1.7)
$$
Therefore, we can obtain the mass relation (1.1) from the assumption
$m_{ei} \propto v_i^2$.
Here, note that although the scalar potential (1.4) is invariant
under the S$_3$ symmetry, but it is not a general one of the 
S$_3$ invariant form. 
As pointed out in Ref.~\cite{Koide06}, the scalar potential with
a general form cannot lead to the relation (1.5).
For the derivation of the VEV relation (1.5), it is essential to
choose the specific form (1.4) of the S$_3$ invariant terms.
Similar formulation is also possible for other discrete symmetries
A$_4$ \cite{Koide0701} and S$_4$ (see below).
However, in such a symmetry, we still need an additional specific
selection rule.
What is the meaning of such a specific selection? 
In the present paper, we investigate this problem by assuming
that the S$_4$ flavor symmetry is embedded into SU(3).

On the other hand, the observed tribimaximal mixing suggests
the following scenario:
From the definition (1.2),
we can denote the fields $(\psi_1, \psi_2, \psi_3)$ as
$$
\left(
\begin{array}{c}
\psi_1 \\
\psi_2 \\
\psi_3
\end{array} \right)
= U_{TB}
\left(
\begin{array}{c}
\psi_\eta \\
\psi_\sigma \\
\psi_\pi 
\end{array}
\right).
\eqno(1.8)
$$
The observed neutrino mixing (1.2) means that when the mass 
eigenstates of the charged leptons are given by the 
$(\psi_1, \psi_2, \psi_3)$ basis, the mass eigenstates of 
the neutrinos are given by the 
$(\psi_\eta, \psi_\sigma, \psi_\pi)$ basis.
Therefore, the problem is to find a model where the charged
lepton mass eigenstates are $(e_1, e_2, e_3)$, while the
neutrino mass eigenstates are given by 
$(\nu_\eta, \nu_\sigma, \nu_\pi)$ with the mass hierarchy
$m_\eta^2 < m_\sigma^2 \ll m_\pi^2$ (or 
$m_\pi^2 \ll m_\eta^2 < m_\sigma^2 $).
In the S$_4$, we can define the same relation as (1.8). 

Thus, the characteristic features (1.1) and (1.2) in the
lepton sector may be understood from the language of S$_4$ 
(also S$_3$ or A$_4$).
However, as seen from the above review, the characteristic features
(1.1) and (1.2) cannot be understood from the S$_4$ symmetry only.
We need some additional assumptions.
In the present model, we will investigate these problems under 
an assumption that
the present S$_4$ symmetry is embedded into an SU(3) symmetry 
\cite{Mohapatra-S4}.
In the next section, the singlet $\phi_\sigma$ and doublet 
$(\phi_\pi, \phi_\eta)$ will be understood as members of a nonet scalar
$\phi$ [{\bf 1}+{\bf 8} of SU(3)], and the VEV relation (1.5) will 
be derived by requiring that $W(\phi)$ is invariant under a Z$_2$
symmetry. 
In Sec.3, in order to give the charged lepton masses and tribimaximal
neutrino mixing, we will discuss the effective Hamiltonian by assuming
an Froggatt-Nelsen \cite{Froggatt} type model.
Finally, Sec.4 will be devoted to the summary and concluding remarks.


\vspace{3mm}

{\large\bf 2 \ VEVs of SU(3) nonet scalars}

The goal in the present section is to obtain the VEV relation (1.6)
[i.e. (1.5)].
As seen in the previous section, in order to obtain the desirable
results (1.5), we need assume an equal weight between 
the doublet and singlet terms of S$_4$.
In the present paper, we assume that the S$_4$ symmetry is embedded into
an SU(3) symmetry.
The doublet $(\phi_\pi, \phi_\eta)$ and singlet $\phi_\sigma$ of
S$_4$ are embedded in the {\bf 6} and $({\bf 8}+{\bf 1})$ of 
SU(3) \cite{Mohapatra-S4}.
In the present model \cite{Koide0705}, we assume that the doublet 
$(\phi_\pi, \phi_\eta)$ 
and singlet $\phi_\sigma$ originate in SU(3) octet and singlet, 
respectively.
The essential assumption in the present paper is that
the fields $\phi_u$ and $\phi_d$ always appear in the theory with the
form of the nonet of U(3):
$$
\phi = \left(
\begin{array}{ccc}
\phi_1^1 & \ast & \ast \\
\ast & \phi_2^2 & \ast \\
\ast & \ast & \phi_3^3 
\end{array} \right) ,
\eqno(2.1)
$$
where
$$
\begin{array}{l}
\phi_1^1 = \frac{1}{\sqrt{3}} \phi_\sigma + \frac{2}{\sqrt{6}} \phi_\eta , \\
\phi_2^2 = \frac{1}{\sqrt{3}} \phi_\sigma - \frac{1}{\sqrt{6}} \phi_\eta 
- \frac{1}{\sqrt{2}} \phi_\pi , \\
\phi_3^3 = \frac{1}{\sqrt{3}} \phi_\sigma - \frac{1}{\sqrt{6}} \phi_\eta 
+ \frac{1}{\sqrt{2}} \phi_\pi , 
\end{array}
\eqno(2.2)
$$
and the index $f$ ($f=u,d$) has been dropped.

Although we have obtained the VEV (1.5) relation from 
the scalar potential (1.4) by calculating $\partial V/\partial \phi_a$
($a=\pi, \eta, \sigma$), in the present paper, we will obtain
the relation (1.5) from an SU(3) invariant superpotential $W$
(The derivation of the VEV relation (1.5) from a superpotential
$W$ has first been attempted by Ma \cite{Ma0612}):
The SU(3) invariant superpotential for the nonet fields $\phi_f$ 
($f=u,d$) are given by
$$
W(\phi_f) = \frac{1}{2} m_f {\rm Tr}(\phi_f\phi_f) + 
\frac{1}{2\sqrt3}\lambda_f {\rm Tr}(\phi_f\phi_f\phi_f).
\eqno(2.3)
$$
Since, in the next section, we want to assign charges +1 and $-1$
of a Z$_3$ symmetry to the fields $\phi_u$ and $\phi_d$, respectively,
we also assign the Z$_3$ charges +1 and $-1$ to the mass parameters
$m_u$ and $m_d$ in Eq.(2.3), respectively.
However, the U(3) invariant superpotential (2.3) cannot give 
the relation (1.5).
As we show below, only when we drop the 
${\rm Tr}[(\phi^{(8)})^3]$-term in the cubic terms
${\rm Tr}(\phi^3)$, we can obtain the VEV relation (1.5).
Therefore, we introduce a Z$_2$ symmetry, and we assign the
Z$_2$ parities $-1$ and $+1$ (the Z$_2$ charge +1 and 0) 
for the octet part $\phi^{(8)}$ and 
singlet part $\phi^{(1)}$ of the nonet field $\phi$, respectively.
The symmetry Z$_2$ breaks U(3) into SU(3).
(In other words, in the present model, the flavor symmetry U(3) 
is explicitly broken from the beginning by the Z$_2$ symmetry. ) 
Under the requirement of the Z$_2$ invariance, i.e.
the invariance under the transformation 
$$
(\phi^{(8)},\phi^{(1)}) \rightarrow (-\phi^{(8)}, +\phi^{(1)}) ,
\eqno(2.4)
$$
the terms ${\rm Tr}(\phi^{(8)}\phi^{(8)} \phi^{(8)})$ 
are forbidden.
Thus, the superpotential (2.3) with the Z$_2$ invariance leads to
$$
W(\phi)=
\frac{1}{2} m \left[ {\rm Tr}(\phi^{(8)}\phi^{(8)})
+\phi_{\sigma}^2 \right]
 + \frac{1}{2} \lambda \phi_{\sigma} \left[
{\rm Tr}(\phi^{(8)}\phi^{(8)}) + \frac{1}{3} \phi_{\sigma}^2
\right] 
$$
$$
=\frac{1}{2}m\left( \phi_{\sigma}^2+ \phi_\pi^2+\phi_\eta^2 \right)
+\frac{1}{2} \lambda \left[( \phi_{\pi}^2+ \phi_\eta^2)\phi_\sigma 
+\frac{1}{3}\phi_\sigma^3 \right] + \cdots .
\eqno(2.5)
$$
From the superpotential (2.5) with the Z$_2$ invariance,
we obtain the VEV relation (1.5) as follows:
From the condition
$$
\frac{\partial W}{\partial (\phi^{(8)})_i^j} = 
m (\phi^{(8)})_j^i + \lambda \phi_{\sigma}
(\phi^{(8)})_j^i =0 ,
\eqno(2.6)
$$
we obtain
$$
m + \lambda \phi_{\sigma} =0,
\eqno(2.7)
$$
for $(\phi^{(8)})_i^j \neq 0$.   
By eliminating $m$ from Eq.(2.7) and the condition 
$$
\frac{\partial W}{\partial \phi_{\sigma}} =
 m \phi_{\sigma}
+\frac{1}{2} \lambda \left[ {\rm Tr}(\phi^{(8)}\phi^{(8)}) 
+ \phi_{\sigma}^2 \right] =0 ,
\eqno(2.8)
$$
we obtain the relation 
$$
\phi_{\sigma}^2 = {\rm Tr}(\phi^{(8)}\phi^{(8)}) =
\phi_{\pi}^2 + \phi_{\eta}^2 + \cdots ,
\eqno(2.9)
$$
where ``$\cdots$" denotes the contributions of ${\bf 3}$ and
${\bf 3}'$ of S$_4$.

The result (2.9) is still not our goal, because the relation contains
the VEVs of the ${\bf 3}$ and ${\bf 3}'$ of S$_4$.
So far, we have not discussed the splitting among the S$_4$ 
multiplets.
Now, we bring a soft symmetry breaking of SU(3) into S$_4$
with an infinitesimal parameter $\varepsilon$ into the mass
term of $W(\phi)$ as
$$
{\rm Tr}(\phi^{(8)}\phi^{(8)}) \Rightarrow 
\phi_{\pi} \phi_{\pi} + \phi_{\eta} \phi_{\eta}
+(1+\varepsilon) \sum_{i\neq j} (\phi^{(8)})_i^j
(\phi^{(8)})_j^i ,
\eqno(2.10)
$$
by hand. 
Recall that when we obtain the relation (2.7), we have assumed
$(\phi^{(8)})_i^j \neq 0$.
Now, the conditions (2.6) are modified into the following conditions:
$$
\left[(1+\varepsilon) m + \lambda \phi_{\sigma} \right]
(\phi^{(8)})_j^i =0 \ \ \ (i\neq j),
\eqno(2.11)
$$
$$
\left(m + \lambda \phi_{\sigma} \right)
\phi_a =0 \ \ \ (a=\pi , \eta),
\eqno(2.12)
$$
Therefore, we must take either $(\phi^{(8)})_j^i =0$ ($i\neq j$) or 
$\phi_a =0$ ($a=\pi , \eta$) for $\varepsilon \neq 0$.  
When we choose the solution
$$
\langle (\phi^{(8)})_j^i \rangle = 0  \ \ \  (i\neq j) ,
\eqno(2.13)
$$
we can obtain the desirable relation (1.5).
(However, it is possible that we can also take another solution
with $\phi_\pi=\phi_\eta=0$ and $(\phi^{(8)})_j^i \neq 0$.
The VEV solutions are not unique.
The result (1.5) is merely one of the possible solutions.)

Thus, we have obtained not only the desirable VEV relation (1.5),
but also the results (2.13).
It should be worthwhile noticing that if we have assume the 
superpotential (2.3) without requiring the Z$_2$ invariance, 
we could obtain neither (1.5) nor (2.13). 

By the way, we know that the three masses in any sectors of 
quarks and leptons are completely different among them. 
Therefore, if we assume a flavor symmetry, 
the symmetry must finally be broken completely.  
Usually, a relation which we derive in the exact symmetry limit 
is only approximately satisfied under the
symmetry breaking.
Although we derive the VEV relation (1.5) under the S$_4$
symmetry, the problem is whether the VEV relation (1.5) which is
obtained under the S$_4$ symmetry 
is spoiled or not when we introduce such a symmetry breaking.
In Ref.\cite{Koide0705}, we will find that such a symmetry
breaking term without spoiling the relation (1.5) is indeed
possible.

\vspace{3mm}

{\large\bf 3 \ Effective Hamiltonian}

If we regard the scalars $\phi_u$ and $\phi_d$ as SU(2)$_L$ doublets, 
such a model with multi-Higgs doublets causes a flavor changing 
neutral current (FCNC) problem. 
Therefore, we must consider that the fields $\phi_u$ and $\phi_d$
are SU(2)$_L$ singlets.
In the present paper, we assume a 
Froggatt-Nielsen \cite{Froggatt} type model
$$
H^{eff}=y_e \overline{\ell}_L H_L^d \frac{\phi_d}{\Lambda}
\frac{\phi_d}{\Lambda} \frac{\xi}{\Lambda} e_R
+y_\nu \overline{\ell}_L H_L^u \frac{\phi_u}{\Lambda}
\frac{\chi}{\Lambda} \nu_R
+y_R \overline{\nu}_R \Phi_R \nu_R^\ast ,
\eqno(3.1)
$$
where $\ell_{iL}$ are SU(2)$_L$ doublet leptons 
$\ell_{iL}=(\nu_{iL}, e_{iL})$, 
$H_L^d$ and $H_L^u$ are conventional SU(2)$_L$ doublet Higgs scalars, 
$\phi_f$ ($f=u,d$), $\xi$ and $\chi$ are SU(2)$_L$ singlet scalars, 
and $\Lambda$ is a scale of the effective theory. 
We consider that $\langle \phi_f \rangle/\Lambda$, 
$\langle \xi \rangle/\Lambda$ and $\langle \chi \rangle/\Lambda$
are of the order of 1. 
The scalar $\Phi_R$ has 
been introduced in order to generate the Majorana mass $M_R$ of the 
right-handed neutrino $\nu_R$.
As we note later, in the present model, the right-handed
neutrinos $\nu_R=(\nu_R^{(+)} +\nu_R^{(-)})/\sqrt2$
are singlets of the SU(3) flavor.
The role of $\xi=(\xi^{(+)}+\xi^{(-)})/\sqrt2$ and $\chi$ will 
be explained later.
In order to understand the appearance of the combinations
$H_L^d \phi_d \phi_d \xi$ and $H_L^u \phi_u \chi$, we assume 
two Z$_3$ symmetries (Z$_3$ and Z$'_3$ in Table 1).
Those quantum number assignments are given in Table 1.
However, even with those quantum numbers, we cannot
distinguish the state $\phi_f^\dagger$ from $\phi_f \phi_f$.
For example, the interaction $\bar{\ell}_L H_d \phi_d^\dagger \xi e_R$
is possible in addition to the interaction 
$\bar{\ell}_L H_d \phi_d \phi_d \xi e_R$.
Although we have started from an SUSY scenario in the previous
section, now, we have adopted an effective Hamiltonian 
which is not renormalizable.
Therefore, in principle,
the interaction $\bar{\ell}_L H_d \phi_d^\dagger \xi e_R$
cannot be ruled out.
For the moment, in order to forbid such an undesirable
term, we assume that the fields which can appear in 
the effective Hamiltonian are confined to holomorphic ones.

\vspace{0.5cm}
\begin{table}
{\bf Table 1} \ SU(3) and S$_4$ assignments of the fields

\vspace{2mm}
\begin{tabular}{|ccccccc|} \hline
Fields & SU(2)$_L$ & SU(3) & S$_4$ & Z$_3$ & Z$_3^{\prime}$ 
& Z$_2$  \\ \hline
$\ell_L$ & {\bf 2}   & {\bf 3}  &  ${\bf 3}'$ & 0  & 0 & 0 \\
$e_R$    & {\bf 1}   & {\bf 3}  &  $ {\bf 3}'$ & 0 & 0 & 0 \\ 
$\nu_R^{(\pm)}$   & {\bf 1}   & {\bf 1} & {\bf 1}& 0 & 0 & 0/+1 \\ \hline
$\phi_u$ & {\bf 1}   &{\bf 1}+{\bf 8}  & $ {\bf 1}+({\bf 2}+{\bf 3}+{\bf 3}')$ 
& +1 & +1 & 0/+1 \\
$\phi_d$ & {\bf 1}   &{\bf 1}+{\bf 8}  & ${\bf 1}+({\bf 2}+{\bf 3}+{\bf 3}')$ 
& $-1$ & $-1$ &  0/+1  \\
$\xi^{(\pm)}$ & {\bf 1}   &{\bf 1}  &  ${\bf 1}$ & 0 & $-1$ & 0/+1 \\
$\chi$ & {\bf 1}   &{\bf 3}  &  ${\bf 3}'$ & +1 & $-1$ & 0 \\
$H_L^u$  & {\bf 2}   &{\bf 1}   & {\bf 1} & +1 & 0 & 0 \\ 
$H_L^d$  & {\bf 2}   &{\bf 1}   & {\bf 1} &  $-1$ & 0 & 0 \\
$\Phi_R$   & {\bf 1}   &{\bf 1} &  {\bf 1} & 0 & 0 &  0 \\\hline
\end{tabular}
\end{table}
\vspace{5mm}


{\bf (a) Charged lepton sector}

Recall that we have already assumed the invariance of the 
superpotential under the Z$_2$ transformation (2.4) in order to drop 
the cubic part of the octet $\phi^{(8)}$.
Therefore, the term $\phi\phi$ means 
$\phi^{(8)}\phi^{(8)} +\phi^{(1)}\phi^{(1)}$ under the Z$_2$ invariance.
However, in order to give $m_{ei} \propto \langle \phi_i^i \rangle^2$,
what we want is not 
$\phi^{(8)}\phi^{(8)} +\phi^{(1)}\phi^{(1)}$,
but $\phi^{(8)}\phi^{(8)} +\phi^{(1)}\phi^{(1)} 
+\phi^{(8)}\phi^{(1)} +\phi^{(1)}\phi^{(8)}$.
In order to evade this problem, we introduce additional 
fields $\xi^{(+)}$ and $\xi^{(-)}$ whose Z$_2$ parity are 
$+1$ and $-1$, respectively.
The effective interactions in the charged lepton sector are given by
$$
H_e^{eff} = \frac{y_e}{\sqrt2} \bar{e}_L^i (\phi_d)_i^j (\phi_d)_j^k
(\xi^{(+)} +\xi^{(-)})  e_{Rk} ,
\eqno(3.2)
$$
where we have dropped the Higgs scalar $H_L^d$ since we discuss
flavor structure only.
The expression (3.2) becomes
$$
H_e^{eff} = \frac{y_e}{\sqrt2}  
\bar{e}_L [ (\phi_d^{(8)}\phi_d^{(8)}
+ \phi_d^{(1)}\phi_d^{(1)})\xi^{(+)} +
(\phi_d^{(8)}\phi_d^{(1)}+ \phi_d^{(1)}\phi_d^{(8)}) \xi^{(-)}]  e_{R}.
\eqno(3.3) 
$$
Since we have assumed that $\xi^{(+)}$ and $\xi^{(-)}$ appear
symmetrically in the theory, we also assume
$$
\langle\xi^{(+)}\rangle=\langle\xi^{(-)}\rangle \equiv v_\xi .
\eqno(3.4)
$$
Then, we obtain the effective Hamiltonian for the charged leptons
$$
H_e^{eff} = \frac{y_e v_d v_\xi }{\sqrt2 \Lambda^3}
\sum_i \bar{e}_L^i \langle (\phi_d^{(8+1)})_i^i\rangle^2 e_{Ri} ,
\eqno(3.5)
$$
where $v_d =\langle H_L^{d0}\rangle$.
Since the fields $(\phi_d)_i^i$ are defined by Eq.(2.2), 
we can obtain the charged lepton mass relation (1.1) 
from the VEV relation (1.6).


{\bf (b) Neutrino sector}

In the present model, the right-handed neutrinos $\nu^{(\pm)}$ are
singlets of SU(3).
Therefore, in the neutrino seesaw mass matrix 
$M_\nu = m_L^\nu M_R^{-1} (m_L^\nu)^T$, $M_R$ is a $1\times 1$
matrix and $m_L^\nu$ is a $3\times 1$ matrix.
In order to compensate for the absence of the conventional triplet 
neutrinos $\nu_R$,  a new scalar $\chi$ which is a triplet of SU(3) 
has been introduced.
The neutrino Dirac mass terms are given by the following effective
Hamiltonian
$$
H_{Dirac}^{eff} =y_\nu \frac{v_u}{\Lambda^2} \bar{\nu}_L^i 
\langle (\phi_u)_i^j\rangle \langle\chi_j \rangle (\nu_R^{(+)}
+\nu_R^{(-)}) ,
\eqno(3.6)
$$
where $v_u= \langle H_L^{u0}\rangle$. 
It is likely that the scalar potential $V(\chi)$ for the SU(3)
triplet $\chi$ has a specific VEV solution
$$
\langle \chi_{1}\rangle =\langle \chi_{2}\rangle =
\langle \chi_{3}\rangle \equiv v_\chi .
\eqno(3.7)
$$
When we assume the VEVs (3.7), we obtain 
$$
H_{Dirac}^{eff} =y_\nu  \frac{v_u v_\chi}{\sqrt2 \Lambda^2} 
(\bar{\nu}_\eta\ \bar{\nu}_\sigma \ \bar{\nu}_\pi)_L
\left[
\left( \begin{array}{c}
v_{\eta} \\
0 \\
v_{\pi}
\end{array} \right) \nu_R^{(-)}
+\left( \begin{array}{c}
0 \\
v_{\sigma} \\
0
\end{array} \right) \nu_R^{(+)} \right] ,
\eqno(3.8)
$$
where $v_a = \langle \phi_{ua}\rangle$ ($a=\pi, \eta, \sigma$)
(for convenience, we have dropped the index $u$).
Therefore, we obtain the effective neutrino mass matrix 
on the $(\eta, \sigma, \pi)$ basis,
$$
U_{TB}^T M_\nu U_{TB} \equiv M_\nu^{(\eta\sigma\pi)} = \frac{1}{M_R^{(-)}} 
\left( \begin{array}{ccc}
v_\eta^2 & 0 & v_\pi v_\eta \\
0 & 0 & 0 \\
v_\pi v_\eta & 0 & v_\pi^2 
\end{array} \right) 
+ \frac{1}{M_R^{(+)}} 
\left( \begin{array}{ccc}
0 & 0 & 0 \\
0 & v_\sigma^2 & 0 \\
0 & 0 & 0 
\end{array} \right) ,
\eqno(3.9)
$$
where  $M_R^{(\pm)} = y_R^{(\pm)}
\langle \Phi_R \rangle$, 
 and we have dropped the common factors 
$(y_\nu v_u  v_\chi/\sqrt2 \Lambda^2)^2$.
By the way, the ratio $v_\pi/v_\eta$ cannot be 
determined from the potential (2.6), and the ratio is determined by
a soft S$_4$ symmetry breaking term $W_{SB}$ which has been
discussed in the previous section.
We can choose a solution $v_\pi =0$ in the superpotential 
$W(\phi_u)$ by adjusting the parameter $\beta$ in $W_{SB}$,
differently from the case of $W(\phi_d)$. 
Then, the neutrino mass matrix (3.9) becomes a diagonal form 
$D_\nu =(1/M_R^{(-)}){\rm diag}( v_\eta^2, 0, 0)+
(1/M_R^{(+)}){\rm diag}( 0, v_\sigma^2, 0)$.
Since the mass matrix $M_\nu$ on the $(\nu_1,\nu_2,\nu_3)
=(\nu_e,\nu_\mu,\nu_\tau)$ basis is given by
$$
M_\nu =U_{TB} M_\nu^{(\eta\sigma\pi)} U_{TB}^T =U_{TB} D_\nu U_{TB}^T ,
\eqno(3.10)
$$
we can obtain the tribimaximal mixing 
$$
U_{\nu}=U_{TB} ,
\eqno(3.11)
$$
and the neutrino masses
$$
m_{\nu 1}=k v_\eta^2 , \ \ m_{\nu 2}=k v_\sigma^2 , \ \ 
m_{\nu 3}=0,
\eqno(3.12)
$$ 
for the case of $M_R^{(+)}=M_R^{(-)}\equiv M_R$,
where $k=(y_\nu v_u v_\chi)^2/2M_R \Lambda^4$ and
$(\nu_\eta, \nu_\sigma,\nu_\pi)$ has been renamed 
$(\nu_1,\nu_2,\nu_3)$ according to the conventional naming.

However, since we have taken $v_\pi=0$, the value of $v_\eta$ satisfies 
$v_\eta^2=v_\sigma^2$ from the relation (1.5), so that the result
(3.12) gives $m_{\nu 1}=m_{\nu 2}$.
The observed value \cite{solar} $\Delta m^2_{solar}$ is small, but it is not
zero.  
Therefore, we must consider a small deviation between the first and second
terms in (3.9) (i.e. $M_R^{(+)} \neq M_R^{(-)}$).
Since the value $M_R^{(-)}/M_R^{(+)}$ is free in the present model,
we cannot predict an explicit value of the ratio
$\Delta m^2_{solar}/\Delta m^2_{atm}$.

Since the present model gives an inverse hierarchy of the neutrino
masses, the predicted effective electron neutrino mass
$$
\langle m_{\nu_e}\rangle =\left|\sum_i U_{ei}^2 m_{\nu i}\right|
\simeq |m_{\nu 1}| 
\simeq |m_{\nu 2}| \simeq \sqrt{\Delta m_{atm}^2} 
=5.23^{+0.25}_{-0.40} \times 10^{-2}\, {\rm eV},
\eqno(3.13)
$$
where we have used the value \cite{atm}
$\Delta m_{atm}^2 = 2.74^{+0.44}_{-0.26}\times 10^{-3}\, {\rm eV}^2$.
This value (3.13) is sufficiently sensitive to the next generation
experiments of the neutrinoless double beta decay.

\vspace{3mm}

{\large\bf 4 \ Summary}

In conclusion, on the basis of the S$_4$ symmetry which is embedded
into SU(3), we have investigated a lepton mass model with the
effective Hamiltonian of the Froggatt-Nielsen type (3.1). 
We have assumed that the singlet and doublet of S$_4$ originate
in the singlet and octet of SU(3), and we have obtained the VEV
relation (1.5). In the derivation of the VEV relation (1.5), 
the essential assumptions for the superpotential $W(\phi_f)$ are 
the following two:
(i) the scalar fields $\phi_f$ always appear in terms of the nonet
form (2.1) of U(3); (ii) the superpotential $W(\phi_f)$ is 
invariant under the Z$_2$ transformation (2.4).
Then, we have obtained not only the VEV relation (1.5), but also
$\langle (\phi^{(8)})_i^j \rangle =0$ ($i\neq j$) for the other
components of $\phi^{(8)}$ (i.e. 
$\langle {\bf 3}\rangle =\langle {\bf 3}' \rangle =0$).

In the charged lepton sector, 
we have assumed the Frogatt-Nielsen-type effective
Hamiltonian 
$\bar{e}_L^i (\phi_d)_i^j (\phi_d)_j^k e_{Rj}$.
Since we have obtained the VEV of $\phi_i^j$
$$
\langle \phi_d \rangle = {\rm diag}\left(
\frac{1}{\sqrt3}v_\sigma +\frac{2}{\sqrt6}v_\eta ,
\frac{1}{\sqrt3}v_\sigma -\frac{1}{\sqrt6}v_\eta -\frac{1}{\sqrt2}v_\pi ,
\frac{1}{\sqrt3}v_\sigma -\frac{1}{\sqrt6}v_\eta +\frac{1}{\sqrt2}v_\pi
\right) ,
\eqno(4.1)
$$
we can obtain
$$
m_{e i} \propto \langle (\phi_d)_i^i \rangle^2 \equiv (v_i^i)^2 ,
\eqno(4.2)
$$
so that the charged lepton masses $m_{ei}$ satisfy the relation
(1.1) under the definition (2.2).
Here, we would like to emphasize that the result 
$\langle\phi_i^j\rangle =0$ for $i\neq j$, Eq.(2.13), is essential
to obtain Eq.(4.2) in the Froggatt-Nielsen-type model.
In the Froggatt-Nielsen-type model, if we had considered a triplet scalar
$\phi$, we could obtain the result $m_{ei} \propto v_i^2$, because 
the effective interaction $\bar{e}_L \phi \phi e_R$ gives 
$\bar{e}_{Li} \phi_i \phi_j e_{Rj}$.
On the other hand, if we had adopted a seesaw-type model, by assuming 
a triplet scalar $\phi$ whose effective interaction is given by
$$
H_e =\sum_i\left( \bar{e}_{Li} \phi_i E_{Ri} + \bar{E}_{Li} \phi_i e_{Ri}
+M_E \bar{E}_{Li} E_{Ri} \right) ,
\eqno(4.3)
$$
we could automatically obtain a form
$$
H_L^{eff} = \sum_i \frac{1}{M_E} \bar{e}_{Li} \phi_i^2 e_{Ri} ,
\eqno(4.4)
$$
through a seesaw mechanism \cite{seesaw}.
However, it is not easy to obtain the VEV relation (1.6) for the 
triplet scalar $\phi_i$.
This is the main motive for introducing the nonet (not triplet) scalar
$\phi$.

For the neutrino sector, we have obtained the tribimaximal mixing (1.2)
by introducing an SU(3) triplet scalar $\chi$ and the two 
SU(3) singlet right-handed neutrinos $\nu_R^{(\pm)}$ 
in addition to the nonet scalar $\phi_u$.
In the present model, the right-handed neutrinos $\nu_R^{(\pm)}$ are 
singlets of SU(3), the Majorana neutrino mass
matrices $M_R^{(\pm)}$ have no flavor structure, (i.e. $M_R$
are  $1\time 1$ matrices).
Also note that the neutrino Dirac mass matrix $m_L^\nu$ is
a $3\times 1$ matrix.
This plays an essential role to derivation of the tribimaximal
mixing.
For the neutrino mass spectrum, since the model
gives $m_{\nu 1}=m_{\nu 2}$ in the limit of $M_R^{(+)}= M_R^{(-)}$,
we must consider a small deviation $M_R^{(+)} \neq M_R^{(-)}$.
Since the value of $M_R^{(-)}/ M_R^{(+)}$ is a free parameter 
in the present model, we cannot predict the value 
$\Delta m^2_{solar}/\Delta m^2_{atm}$ at present, although the smallness
of the ratio $\Delta m^2_{solar}/\Delta m^2_{atm}$ can be understood.
Since the present model gives an inverse hierarchy of the neutrino
masses, we can predict the effective electron neutrino mass 
$\langle m_{\nu_e}\rangle \simeq 0.05$ eV,
which is sufficiently sensitive to the next generation
experiments of the neutrinoless double beta decay.

The present model seems to provide suggestive hints on seeking for 
a model which leads to the tribimaximal mixing (1.2) and the charged 
lepton mass relation (1.1), although the model has still many
points which should be improved. 
We summarize some of the future tasks:

\noindent (i) Seeking for a more natural mechanism
which can provide $m_{ei} \propto v_i^2$,
apart from a Froggatt-Nielsen-type model.

\noindent (ii) Seeking for a model which is completely predictable
neutrino mass spectrum (in the present model, $\Delta m_{21}^2$
was not predictable) together with the nearly tribimaximal mixing.

\noindent (iii) In the present model, the right-handed neutrino $\nu_R$  was 
a singlet of SU(3).
However, it is likely that $\nu_R$ still a triplet.

\noindent (iv) Seeking for the origin of the symmetry breaking of
SU(3) into S$_4$.

We hope that the present model will also provide a promising clue to 
the unified mass matrix model of the quarks and leptons.

\vspace{4mm}

\centerline{\large\bf Acknowledgments} 

The author would like to thank E.~Takasugi, H.~Fusaoka 
and N.~Haba for helpful conversations.
This work is supported by the Grant-in-Aid for
Scientific Research, Ministry of Education, Science and 
Culture, Japan (No.18540284).



\end{document}